\documentstyle[prl,aps,epsf]{revtex}
\begin{document}

\advance\textheight by 0.2in
\draft
\twocolumn[\hsize\textwidth\columnwidth\hsize\csname@twocolumnfalse
\endcsname  

\title{Glassiness Vs. Order in Densely Frustrated Josephson Arrays}

\author{Pramod Gupta,$^a$ S. Teitel,$^a$ and Michel J. P. Gingras$^b$}
\address{$^a$Department of Physics and Astronomy, University of Rochester, 
Rochester, New York 14627}
\address{$^b$Department of Physics, University of Waterloo, Waterloo, 
Ontario N2L-3G1, Canada}
\date{\today}
\maketitle

\begin{abstract}
We carry out extensive Monte Carlo simulations of the Coulomb gas
dual to the uniformly frustrated two dimensional XY model,
for a sequence of frustrations $f$ converging to the irrational
$(3-\sqrt{5})/2$.  We find in these systems a sharp
first order equilibrium phase transition to an ordered vortex
structure at a $T_c$ which varies only slightly with $f$.  
This ordered vortex structure remains 
in general phase incoherent until a lower vortex pinning
transition $T_p(f)$ that varies with $f$.
We argue that the glassy behaviors reported for this
model in earlier simulations are dynamic effects.
\end{abstract}


]

The glass transition to a frozen disordered state remains one
of the oldest unresolved problems of condensed matter physics.
While much progress had been made in the area of ``spin
glasses'', in which the glassy state is a consequence of
intrinsic random frustration in the Hamiltonian, the problem of
``structural glasses'', which possess no intrinsic random disorder,
remains poorly understood \cite{R2,R2.5}.
It is therefore desirable to search for glasslike transitions
in simple intrinsically disorder-free statistical models.
One such candidate system is the two dimensional (2D)
{\it uniformly frustrated} XY model, which models
a periodic array of Josephson junctions in a transverse
applied magnetic field \cite{R3,R3.5}.  Varying the frustration parameter $f$
(magnetic field) of this model leads to complex commensurability 
effects between the underlying discrete grid and the vortex 
lattice that forms in response to the frustration \cite{R4}.  
Some years ago, 
Halsey \cite{R5} presented numerical evidence that, in the limit of an 
irrational $f^*=(3-\sqrt{5})/2$, this model displays a finite 
temperature glass 
transition $T_g$ to a superconducting frozen disordered vortex state 
\cite{R6}.  Experiments on superconducting 
wire networks with Halsey's irrational 
$f^*$ have found evidence for a finite $T_g$ from 
the scaling of current-voltage
(IV) characteristics \cite{R6.5}.  However simulations by Granato \cite{R7}, 
using resistively shunted junction dynamics, found an IV
scaling consistent
with $T_g=0$.  Recently, Kim and Lee \cite{R8} have re-investigated 
this problem using Langevin simulations.  They find that near
Halsey's $T_g$, the system's dynamics resembles 
the primary relaxation of supercooled liquids rather than 
that of a spin glass. 

In view of the above conflicting results, 
it is important to establish the true equilibrium
behavior of this model.  We therefore
re-investigate Halsey's problem by carrying out
Monte Carlo (MC) simulations of the 2D Coulomb gas which
is dual to the uniformly frustrated XY models.  
Working with vortex variables directly allows us greater
control over the most relevant slow variables involved in
equilibration, as compared to using the phase variables
of the original XY model \cite{R5,R6,R7,R8}.  
Following Halsey, we consider the
frustrations $f=5/13$, $8/21$, $13/34$, and $21/55$,
which are the first few members of a Fibonacci sequence of 
rational approximants which converges to the irrational 
$f^*=(3-\sqrt{5})/2$.  We find that the low temperature state that
is reached upon slow cooling is highly sensitive to both
the dynamics used as well as the system length $L$.
The true ground states for such $f=p/q$ are much more
complex than previously believed, even for relatively small values
of $q$.  We find that when next-nearest-neighbor vortex hops are included
in the dynamics, all cases show clear evidence for
a sharp {\it first order} equilibrium phase transition $T_c$ near Halsey's
$T_g$, to an $ordered$ vortex structure
consisting of completely filled, completely empty, and partially
filled diagonals.  Below $T_c$, vortices in the partially filled
diagonals can remain mobile, destroying phase coherence in the
direction transverse to the diagonal.  
These vortices pin to the grid, leading
to phase coherence in all directions, only at a lower $T_p(f)$ that
varies with $f$.  We therefore conclude that the ``glass
transition'' observed by Halsey, and the supercooling observed
by Kim and Lee,
is a consequence of energy barriers in their particular dynamics 
inhibiting what is a true first order equilibrium phase transition
to a non-glassy ordered state.  

The uniformly frustrated 2D XY model, within the Villain 
approximation \cite{R9},
can be mapped \cite{R10} onto the following Hamiltonian for a one component 
Coulomb gas on a neutralizing background,
\begin{equation}
{\cal H}_{\rm CG}[n_i]={1 \over 2} \sum_{i,j} (n_{i} - f)G'({\bf r}_{i}-{\bf 
r}_{j})(n_{j} - f) \enspace.
\label{eH}
\end{equation}
The sum is over all sites of a 2D periodic square $L\times L$ grid.
$n_i$ is the integer charge on site $i$, representing
the vorticity of the XY phase angle.  The frustration parameter
$f$, acting as a background charge density, represents the density 
of flux quanta in the applied magnetic
field.  The interaction is $G'({\bf r})=G({\bf 
r})-G({\bf 0})$, where $G({\bf r})$ is the lattice Coulomb 
potential in 2D, with periodic boundary conditions.  
Charge neutrality, $\sum_in_i=L^2f$, is imposed.  See Ref.\,\cite{R11}
for further details.

The elementary move of our MC procedure is the insertion of
a randomly positioned vortex-antivortex pair, $\Delta n_i=+1$, 
$\Delta n_j=-1$, which is then either accepted or rejected by a 
standard Metropolis algorithm.  When we restrict $i$ and $j$ to
nearest-neighbor sites, we find glassy results qualitatively similar
to Halsey's.  When we allow $i$ and $j$ to include 
next-nearest-neighbor sites as well, we find equilibration at low 
temperatures to be dramatically improved.  Our results below are
for this latter dynamics.  Simulations were carried out
cooling from an initial random configuration.  At each temperature
20,000 initial MC passes are discarded for equilibration, with an additional 
1,280,000 MC passes for computing averages. One MC pass refers to 
$L^2$ elementary moves.

\begin{figure}
\epsfxsize=3.0truein
\epsfbox{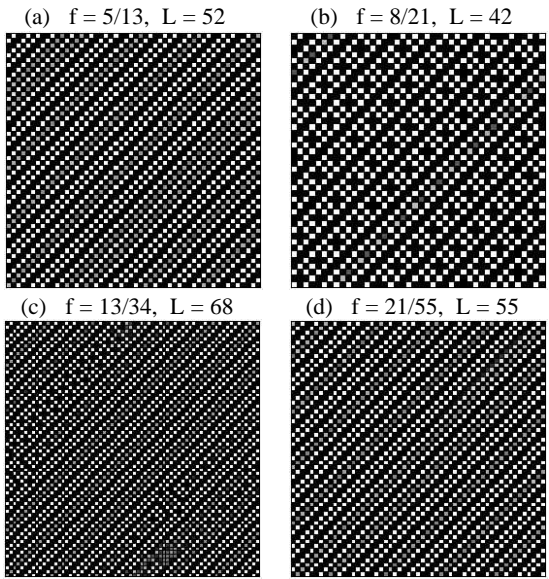}
\caption{Intensity plot of ordered vortex states at $T=0.02<T_c$.
White squares contain vortices, black squares contain no vortices,
and gray squares have an average vortex occupation of $0<\langle n_i
\rangle <1$.}
\label{f1}
\end{figure}

In Fig.\,\ref{f1} we show intensity plots for the {\it average}
vorticity at each site, at $T=0.02<T_c\simeq 0.03$.
Black squares are sites with $\langle n_i\rangle 
\simeq 0$; white squares are sites with
$\langle n_i\rangle \simeq 1$; gray squares are sites with an
average vortex occupation of $0<\langle n_i\rangle <1$.
We find an ordered 
sequence of completely filled, completely empty, and partially filled
diagonals, clearly different from the disordered structures found
by Halsey.  For Figs.\,\ref{f1}a,c,d, we find
translational invariance along the diagonals, except for
occasional defects.  For $f=8/21$ 
(Fig.\,\ref{f1}b) the partially filled diagonals have a pinned
vacancy on every third site.  
We do not know that the
states shown in Fig.\,\ref{f1} represent the true ordered states 
in the thermodynamic limit.  
For $f=5/13$, for example, using
$L=26$ and $L=52$ resulted in a differing sequence for the
filled, empty, and partially filled diagonals.  For $f=8/21$,
the vortices in the partially filled diagonals occupy exactly $2/3$
of the sites in these diagonals.  We may speculate that in the 
true ground state these vortices will from a periodic lattice
with the same structure as the $f=2/3$ ground state \cite{R3}.  
Such a structure can only be made perfectly periodic,
and commensurate with the periodicity of the 
diagonals, when $L$ is an integer multiple of $84$.  The
structure shown in Fig.\,\ref{f1}b, with $L=42$, consists of
such an $f=2/3$--like arrangement, however with
a domain wall introduced by our choice of a too small
value of $L$.  From such considerations we conclude that the
true ground state, for all but the simplest of $f=p/q$,
may involve rather subtle and {\it a priori} unknown
commensurability requirements; its
square unit cell will be of length $mq$ where $m$ may well
be an integer several times larger than previously believed \cite{R12}.
\begin{figure}
\epsfxsize=3.3truein
\epsfbox{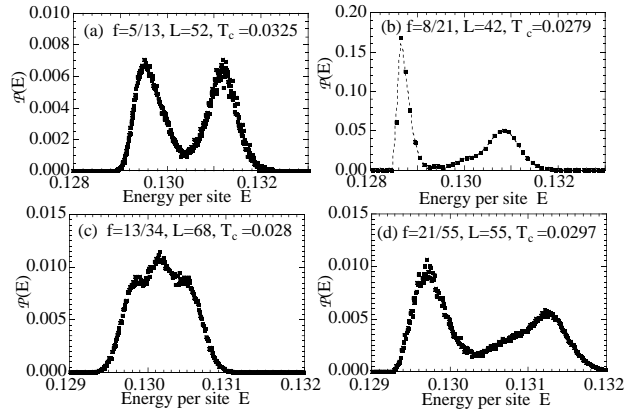}
\vskip .1truein
\caption{
Bimodal energy distribution ${\cal P}(E)$ at $T_c$. 
The first order transition temperature
$T_c\simeq 0.03$ is fixed by the condition that the peaks subtend equal areas.
}
\label{f2}
\end{figure}
For each case we find that the transition to the ordered state is
sharp and seemingly first order.  To demonstrate this, we compute
the histogram of energies ${\cal P}(E)$ encountered at each 
value of temperature
in the simulation.  In both high and low temperature regions,
this histogram is unimodal.  However in a narrow temperature
range when the ordered state first appears, the distribution
becomes bimodal corresponding to the two coexisting states at a
first order transition.  Using an extrapolation technique \cite{R13} we
determine $T_c$ as the temperature for which the
two peaks of the bimodal histogram subtend equal $area$ \cite{R14}.  
Our results are shown in Fig.\,\ref{f2}.
For Figs.\,\ref{f2}a,b,d, the two peaks are 
well separated; the less clear case of $f=13/34$ is perhaps a 
reflection of
the larger concentration of defects seen in Fig.\,\ref{f1}c, or
the possibility that our size $L=68$ still gives too
poor an approximation of the true ground state.  In the 
mapping from the XY model to the
Coulomb gas the temperature has been rescaled \cite{R11} so that
$T^{\rm XY}=2\pi T^{\rm CG}$.  Our value $T^{\rm CG}_c\simeq 0.03$
is thus reasonably close to Halsey's value of $T_g^{\rm XY}\simeq
0.25$.  

Next we consider the superconducting phase coherence
of the Josephson array.  In the original uniformly
frustrated XY model of phase angles $\theta_i$, 
the issue of phase coherence may be
addressed by considering the dependence of the total free 
energy ${\cal F}$ on the net phase angle twist $\Delta_\mu$ that is applied 
across the sample as a boundary condition, i.e. $\theta({\bf r}_i+
L\hat\mu)=\theta({\bf r}_i)+\Delta_\mu$.  If ${\cal F}[\Delta_x,
\Delta_y]$ is independent of the $\Delta_\mu$, then phase coherence
is lost.  Doing the duality transformation to the Coulomb gas
carefully \cite{R10,R15,R16}, one finds that such a fixed twist 
boundary condition
results in an additional term to the Coulomb gas Hamiltonian of 
Eq.\,(\ref{eH}),
\begin{equation}
  \delta{\cal H}[{\bf p};\Delta_x,\Delta_y]=
                 V({2\pi p_x\over L}-\Delta_y)+
                 V({2\pi p_y\over L}+\Delta_x)
\label{edH}
\end{equation}
where
\begin{equation}
   {\bf p}=\sum_i {\bf r}_i n_i
\label{ep}
\end{equation}
is the total ``dipole moment'' of the vortices, and
\begin{equation}
  V(\phi)= -T\ln\left[\sum_{m=-\infty}^\infty e^{-{1\over 4\pi T}
           (\phi-2\pi m)^2}\right]
\label{eV}
\end{equation}
is the Villain function \cite{R9} with coupling $2\pi T$.
The partition function 
for the system with a fixed twist $\Delta_\mu$
is then $Z[\Delta_x,\Delta_y]=Z_{\rm CG}
\langle e^{-\delta{\cal H}[{\bf p};\Delta_x,\Delta_y]/T}\rangle$. 
Here $Z_{\rm CG}$ is the partition function for the 
ensemble defined by 
${\cal H}_{\rm CG}$ of Eq.\,(\ref{eH}) alone, and the average
is with respect to this ensemble (${\cal H}_{\rm CG}$ can be
considered as the ensemble in which $\Delta_\mu$ is averaged over,
and so one has ``fluctuating twist boundary conditions'' in the XY model
\cite{R15}).  
The total free energy is then ${\cal F}[\Delta_x,\Delta_y]
={\cal F}_{\rm CG}+\delta{\cal F}[\Delta_x,\Delta_y]$, where
${\cal F}_{\rm CG}=-T\ln Z_{\rm CG}$ and
\begin{equation}
  \delta{\cal F}[\Delta_x,\Delta_y]
      = -T\ln \sum_{p_x,p_y}{\cal P}(p_x,p_y)e^{-\delta{\cal H}[{\bf p}
      ;\Delta_x,\Delta_y]/T}
\label{eF}
\end{equation}
where ${\cal P}(p_x,p_y)$ is the histogram of total dipole moments
${\bf p}$ at a given temperature, found in the simulation 
using ${\cal H}_{\rm CG}$ of Eq.\,(\ref{eH}).
By storing this 2D histogram, we can therefore deduce the dependence
of the free energy on $all$ values of applied twist $\Delta_\mu$.

\begin{figure}
\epsfxsize=3.0truein
\epsfbox{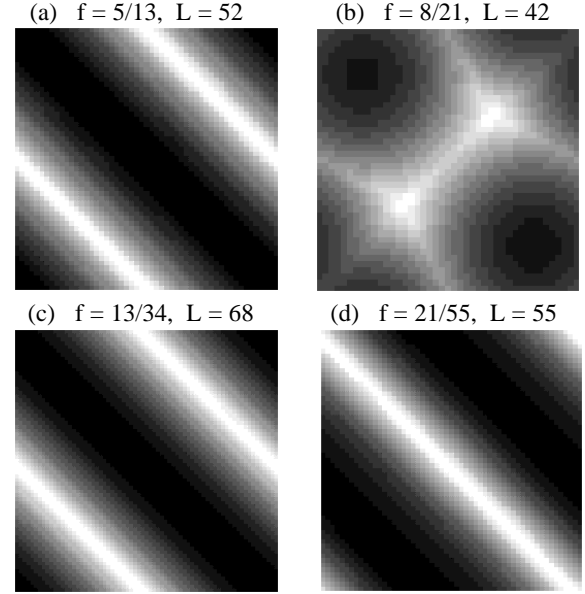}
\vskip .1truein
\caption{Intensity plot of total
free energy $\delta{\cal F}[\Delta_{x},\Delta_{y}]$
at $T=0.02<T_c$. 
Black (white) denotes the functional minima (maxima). 
}
\label{f3}
\end{figure}

In Fig.\,\ref{f3} we show intensity plots of $\delta{\cal F}[\Delta_x,
\Delta_y]$ for $\Delta_\mu\in (-\pi,\pi)$ at $T=0.02<T_c$,
corresponding to the real space plots of Fig.\,\ref{f1}.
For $f=8/21$, Fig.\,\ref{f3}b, we see a rotationally symmetric
parabolic minimum indicating that the system is phase coherent
in all directions (that there are actually two such minima, 
is a result of the thermal motion of the domain wall inserted
by our choice of $L=42$).  
For the other cases however, we see that
while $\delta{\cal F}$ has a parabolic minimum along the direction parallel
to the ordered diagonals, it is constant for the direction 
perpendicular to these diagonals.  This indicates that the vortices
in the partially filled diagonals are free to move along these
diagonals and so destroy phase coherence transverse to this
direction.  Noting that an applied electric current exerts
a force on the vortices which is transverse to the direction
of the current, 
we would expect the Josephson array to have a finite linear
resistivity for all cases except when the current is applied
exactly parallel to the partially filled diagonals.  Thus
the structural transition at $T_c$ is not in general the
superconducting transition of the array.

\begin{figure}
\epsfxsize=3.3truein
\epsfbox{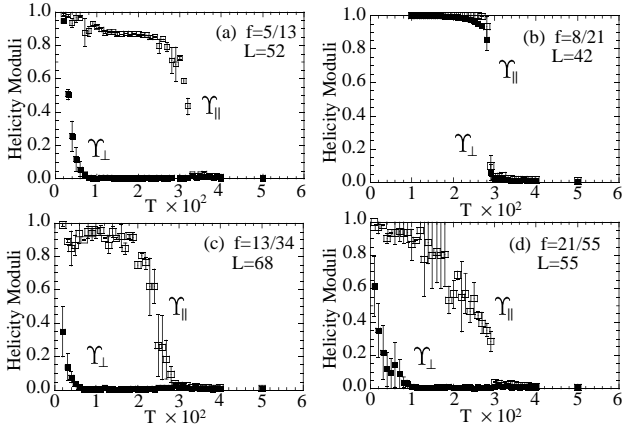}
\vskip .1truein
\caption{Helicity modulus eigenvalues $\Upsilon_\parallel$ and
$\Upsilon_\perp$ vs. $T$. 
$\Upsilon_\parallel$ is non-zero below $T_c\simeq 0.03$, however
$\Upsilon_\perp$ is non-zero only below $T_p(f)\le T_c$.
}
\label{f4}
\end{figure}

The helicity modulus tensor $\Upsilon_{\mu\nu}$
of the 2D XY model is defined by \cite{R17}
\begin{equation}
  \Upsilon_{\mu\nu}\equiv {\partial^2{\cal F}\over
  \partial\Delta_\mu\partial\Delta_\nu}
\label{eUeps}
\end{equation}
where the derivatives are evaluated at the value $\Delta_{\mu 0}$
that minimizes ${\cal F}$.  
From Fig.\,\ref{f3} we expect that $\Upsilon_{\mu\nu}$
is diagonal in a basis that is aligned with the grid diagonal
directions.  We therefore denote by $\Upsilon_\perp$ and
$\Upsilon_\parallel$ the eigenvalues of $\Upsilon_{\mu\nu}$
in the directions perpendicular and parallel to the ordered diagonals 
respectively.  In Fig.\,\ref{f4} we plot $\Upsilon_\perp$
and $\Upsilon_\parallel$ as functions of $T$, for the same values
of $f$ and $L$ as in Figs.\,\ref{f1} and \ref{f3}.  As expected
from Fig.\,\ref{f3} we see that
for all cases $\Upsilon_\parallel$ increases from zero as $T$ 
decreases below $T_c$.  However, except for the case $f=8/21$,
$\Upsilon_\perp$ remains zero below $T_c$, becoming non-zero
only at a lower temperature $T_p(f)$ when the vortices in the
partially filled diagonals pin to the grid.  Similar behavior,
of mobile vortex ``defects'' in an otherwise ordered
vortex structure,
has been observed previously \cite{R19} in simulations of the $f=5/11$ model.
If the vortices in the partially filled diagonals remain mutually
correlated, the region $T_p(f)<T<T_c$ can be described as
a ``floating smectic'' phase, as first postulated by Ostlund \cite{R20}.
If however the correlations are short ranged, as in a 
liquid, one might imagine that vortex hopping between the
different partially filled diagonals may also be possible.
In this case, our result that
$\Upsilon_\parallel>0$ for $T_p(f)<T<T_c$ might be a reflection
of the very high energy barrier for such inter-diagonal hops,
rather than a true phase coherence effect.

We conclude that the sequence of rational $f$ that approach
the irrational $f^*=(3-\sqrt{5})/2$ all undergo a first order
equilibrium phase transition to an ordered vortex structure
at a $T_c\simeq 0.03$.  The exact sequence of the filled, empty,
and partially filled diagonals in this ordered structure 
remains in general unknown for the true ground state in the
thermodynamic limit, however in the cases when we varied $L=mq$ 
for fixed $f=p/q$, we found that $T_c$ remained approximately $0.03$.
The transition to the true superconducting state, with phase
coherence in all directions, occurs in general at a lower 
$T_p(f)$, which can show considerable variation with the frustration $f$.
While this is in qualitative agreement with arguments by Teitel
and Jayaprakash \cite{R3}, which suggested that the superconducting transition
temperature would be a very discontinuous function of $f$,
we as yet can discern no systematic dependence on $f=p/q$
nor can we be certain that the values of $T_p(f)$ obtained here
will not vary if one increases the system size $L$.

Our equilibrium transition at $T_c$ to an ordered vortex
structure was only obtained when we included next-nearest-neighbor
hops in our vortex dynamics.  When moves were
restricted to nearest-neighbor hops only, our simulations
fell out of equilibrium into a frozen disordered state below
$T\simeq 0.033$.  Fig.\,\ref{f1} suggests
why this is so.  Next-nearest-neighbor hops
allow vortices to travel directly up and down the partially filled 
diagonals.  To make such a move using only nearest-neighbor hops, 
one must first hop to a neighboring diagonal, putting three vortices
mutually adjacent.  We find for the energy barrier of
such moves $\Delta E\simeq 0.23-0.35$, depending on $f$, 
and so these moves tend to freeze
out by the temperature $T_c\ll\Delta E$.  Indeed, when 
restricting to nearest-neighbor hops only, we found glassy
behavior even for the simple frustration $f=3/8$ at large $L$,
unless very slow and careful cooling was used.  
Such glassy behavior is therefore more a reflection of the
frustration being non-trivially dense rather than specifically
irrational.

Our results indicate
that the ``glass transition'' observed by Halsey \cite{R5} is an
artifact of his choice of dynamics. 
Our observation of an equilibrium first order transition
strengthens the analogy to structural glasses and
gives a natural explanation for the supercooled relaxation
observed by Kim and Lee, whose simulations were carried out
for the parameters $f=13/34$, $L=34$.
Further work is required to investigate
whether such a supercooled
state can have a well defined finite temperature
glass transition below $T_c$,
or whether, as suggested by Granato, this glass transition
is at $T=0$.

We thank Prof. Y. Shapir for interesting 
discussions.
This work has been supported by U.S. Department of Energy grant 
DE-FG02-89ER14017.





\end{document}